\documentclass[aps,prl,twocolumn,superscriptaddress,nofootinbib,10pt]{revtex4-2}

\usepackage{graphicx} 
\usepackage{dcolumn} 
\usepackage{bm} 
\usepackage{slashed}
\usepackage{xcolor}
\usepackage{amsmath}
\usepackage{lmodern}
\usepackage{dsfont}
\usepackage{hyperref}

\usepackage[letterpaper,top=1.0in, inner=0.5in, outer=0.5in,bottom=1.0in,,headheight=0pt,headsep=20pt,centering]{geometry}

\newcommand{\cannonball}{\textsc{cannonball}}

\begin{document}

\title{Solving Combinatorial Problems at Particle Colliders Using Machine Learning}

\author{Anthony Badea} 
\email{abadea@g.harvard.edu}
\affiliation{Harvard University, Cambridge, Massachusetts, 02138}%

\author{William James Fawcett} 
\affiliation{Cavendish Laboratory, University of Cambridge, Cambridge, CB3 0HE, UK}

\author{John Huth} 
\affiliation{Harvard University, Cambridge, Massachusetts, 02138}%

\author{Teng Jian Khoo} 
\affiliation{Humboldt-Universit\"at zu Berlin, Berlin, 12489, Germany}%

\author{Riccardo Poggi} 
\affiliation{Universit\'e de Gen\`eve, Geneva, 1211, Switzerland}

\author{Lawrence Lee} 
\email{lawrence.lee.jr@cern.ch}
\affiliation{Harvard University, Cambridge, Massachusetts, 02138}%
\affiliation{University of Tennessee, Knoxville, Tennessee, 37996}

\date{\today}

\begin{abstract}
High-multiplicity signatures at particle colliders can arise in Standard Model processes and beyond. With such signatures, difficulties often arise from the large dimensionality of the kinematic space. For final states containing a single type of particle signature, this results in a combinatorial problem that hides underlying kinematic information. We explore using a neural network that includes a Lorentz Layer to extract high-dimensional correlations. We use the case of squark decays in $R$-Parity-violating Supersymmetry as a benchmark, comparing the performance to that of classical methods. With this approach, we demonstrate significant improvement over traditional methods.
\end{abstract}

\maketitle

At particle colliders such as the Large Hadron Collider (LHC)~\cite{Evans_2008}, there is a need to probe high-multiplicity signatures to study the decays of Standard Model (SM) particles and to search for new physics. There, the visible decay products from proton-proton interactions are measured by detectors, such as CMS and ATLAS, as a collection of $N$ four-momentum vectors. Since each momentum contains four components, the measured kinematic feature space per reconstructed collision event is $\mathds{R}^{4N}$. When the decay chains are complex and $N$ is large, this dimensionality can be large enough that combinatorics prevent effective interpretation of the decay history. The aim of this letter is to highlight the capacity of machine learning (ML) to handle combinatorial problems more accurately and robustly than existing methods. 

We are interested in event \textit{interpretation}: the problem of matching parent and child particles. One often has many candidate interpretations of the decay history and must decide which  is correct or accept a background from the incorrect interpretations. In the limit that all final-state particles have the same kind of experimental signature, even events with low multiplicity are swamped with combinatorial background making explicit mass resonance reconstruction or crafting useful high-level observables difficult.

Since this problem has existed for decades, classical techniques (\emph{i.e.} without ML) have been developed to overcome it. Minimizations over the set of combinations in invariant mass or angle are particularly popular for simple signatures \cite{ATLAS:2019npw}. For more complex scenarios, other analyses have used event-level $\chi^2$ minimization \cite{D0:1995jca}, the Dalitz variables for three-body decays \cite{PhysRev.94.1046,CMS:2018ikp}, jet substructure \cite{CMS:2018pdq, ATLAS:2012dp}, and accidental substructure for avoiding combinatorics \cite{Cohen:2012yc,CMS:2012zlv,ATLAS:2018umm}. 

In this work, we demonstrate how a simple ML model can outperform the two most prominent existing methods for an example benchmark. We focus on a case inspired by $R$-parity-violating (RPV) Supersymmetry (SUSY) as a proof of concept \cite{Barbier:2004ez}, with a simplified model where pair-produced heavy stop squarks $\tilde{t}$ decay to two SM quarks each, giving a four-jet final state at leading order. This signature is ideal for our study because the all-hadronic $(2\times2)$ jet topology is one of the simplest cases facing this combinatorial background. Its relative simplicity allows traditional methods to retain some utility, unlike in more complex cases. Despite this simplicity, however, even this signature remains under-probed at the LHC ~\cite{Evans:2012bf,ATLAS:2017jnp,CMS:2018mts}.

For the $(2\times2)$ topology, at leading order, there are $\binom{4}{2}/2=3$ possible unique pairings. As only one of these is correct, a simple brute-force approach faces a 200\% combinatorial background, even in a pure sample of signal events. The problem becomes even more difficult in practice due to the possible existence of additional high-momentum jets from initial state radiation (ISR) and overlapping particle interactions (pileup). This background is on top of fully-SM background processes with multiple jets (primarily from pure QCD processes).

\begin{figure}[tb]
    \centering
    \includegraphics[width=8.6 cm]{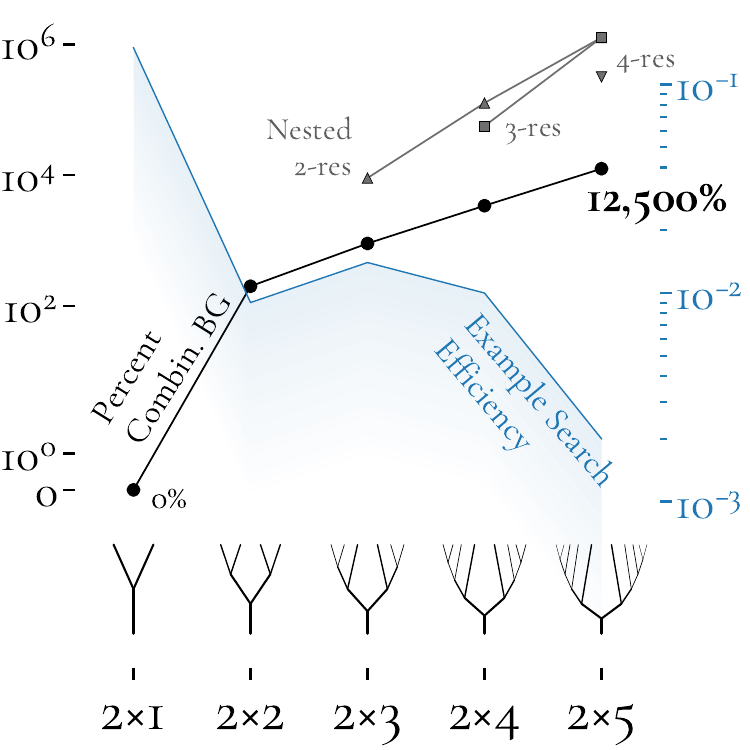}
    \caption{Percent combinatorial background given a brute-force analysis (black) is shown as a function of final-state multiplicity. Increased complexities from nested resonances are shown for various intermediate decays (grey). Example past search signal selection efficiencies are also shown (blue) for $1$~TeV resonances from recent LHC searches~\protect\cite{CMS:2018pdq,CMS:2018mts,CMS:2018ikp,ATLAS:2019fgd,ATLAS:2017jnp,ATLAS:2015xmt,ATLAS:2018umm}.
    }
    \label{fig:multijetComb}
\end{figure}

Figure~\ref{fig:multijetComb} shows the level of combinatorial background as a function of the complexity of symmetrically decaying pair-produced multijet resonances. For larger multiplicities, brute-force solutions are significantly less effective. These high multiplicities may also arise from multi-step cascade decays via on-shell intermediate resonances. In the RPV model, $(2\times4)$ and $(2\times5)$ signatures can easily arise when the RPV coupling is smaller than the gauge couplings and a cascade of decays to lighter BSM particles can occur. These more complex topologies are particularly interesting because they can contain nested resonance structure. For example, simple RPV models may yield gluino $\tilde{g}$ pair production with each $\tilde{g}$ decaying to $qq\tilde\chi$, and the $\tilde\chi$ further decaying to $qqq$. To correctly interpret this topology, one must overcome the $\binom{10}{5} / 2 = 126$ unique combinations to reconstruct each 5-jet resonance $\tilde{g}$, and then the additional $\binom{5}{3}=10$ possibilities to resolve \textit{each} 3-jet resonance $\tilde{\chi}$ correctly. The fully correct interpretation of the event is swamped in a combinatorial background of more than a million percent. Past searches for these signatures have proven rather weak for these reasons, as indicated in terms of some past search signal selection efficiencies shown in blue in Figure~\ref{fig:multijetComb}.~\cite{CMS:2018pdq,CMS:2018mts,CMS:2018ikp,ATLAS:2019fgd,ATLAS:2017jnp,ATLAS:2015xmt,ATLAS:2018umm,Evans:2014gfa}

Similar challenges exist in SM measurements~\cite{Alhazmi:2022qbf,Qiu:2022xvr}. However, in searches for new particles, neither the location nor the width of a mass peak is known a priori. Therefore, a method must perform well across a broad range of potential signal masses and lineshapes to be useful in BSM searches. Combinatorial problems in SM event reconstruction are at a significant advantage of being able to use expected resonance masses as strong additional constraints. In the new particle searches discussed here, only aggregate peaking qualities and event-level symmetries can be exploited to reduce the complexity.

Classically, the problem can be constrained significantly beyond the brute-force approach by optimizing over some observable quantity. One prominent example for pair-production is the asymmetry variable

\begin{eqnarray}
    \mathcal{A} = \frac{|m_1-m_2|}{m_1+m_2},
\end{eqnarray}

\noindent where $m_1$ and $m_2$ represent the invariant masses of some exclusive subsets of objects. In this method, the combinatorial solution is taken as the particular configuration of objects into systems 1 and 2 that minimizes $\mathcal{A}$. This technique generally works well since at leading order the invariant mass of half of the objects should be equal to that of the other half. However, the performance drops with realistic detector resolutions since the width of the invariant mass distribution can allow for correct configurations with large $\mathcal{A}$.

Another approach is to minimize the opening angle between grouped objects using

\begin{eqnarray}
    \Delta R^{\Sigma} = \sum_{i\in \{1,2\}} | \Delta R_{i} - C |,
\end{eqnarray}

\noindent where $\Delta R_i \equiv \sqrt{\Delta\eta_i^2+\Delta\phi_i^2}$ is the opening angle between momenta in group $i$ with pseudorapidity $\eta$ and azimuthal angle $\phi$, and $C$ is a customizable offset to optimize for nonzero opening angles that may be favored by the target signal. This method is successful when the resonances are produced well above threshold, with large back-to-back transverse boost. Using the limiting relation that the scale of the opening angle between the decay products of each resonance is roughly $ \left(2m/p_{\mathrm{T}}\right)$ for a parent particle of mass $m$ and transverse momentum $p_{\mathrm{T}}$, highly-boosted resonances tend to have collimated decay products such that $\Delta R^{\Sigma}$ minimization performs well. This has been the primary method (with $C\approx1$) used by ATLAS and CMS for all-hadronic RPV $\tilde{t}$ searches \cite{ATLAS:2017jnp,CMS:2018mts}. 

While these techniques have been useful, there is a significant amount of kinematic information not utilized, motivating the use of ML for this problem. Convolutional Neural Networks have been used on event images \cite{Bhimji:2017qvb,Guo:2018hbv}, but are not ideal because using images requires a choice of pixel segmentation. This results in sparse and higher dimensional event representations while neglecting much of the complex jet physics that produces well-understood four-vectors. For example, an image of size $50\times50$ with three channels such as the tracker, ECAL, and HCAL has $7,500$ dimensions, a far cry from the $16$ dimensions of the original $(2\times2)$ problem. More recently, there is ongoing work on architecture properties tailored to kinematics problems, as in attention-based networks, for example~\cite{Shmakov:2021qdz,Fenton:2020woz,Lee:2020qil,CMS:2021beq}. Recent studies also explore the use of quantum annealers to solve similar event topology problems~\cite{Kim:2021wrr}.

\begin{figure}[t!]
    \centering
    \includegraphics[width=8.6 cm]{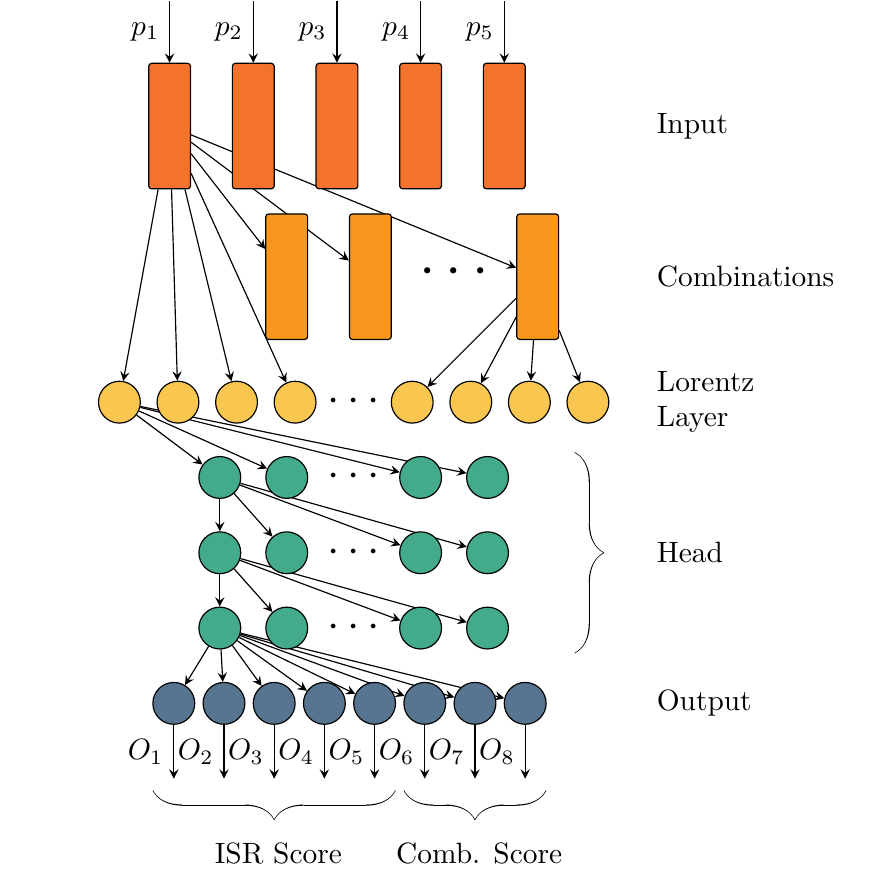}
    \caption{Network structure of \cannonball\ including combination and Lorentz Layers. Only some connections are shown for clarity. All neurons from the Lorentz Layer and after are fully connected. Neurons represented by a circle correspond to scalar values, while rectangles denote four-vector-valued neurons.}
    \label{fig:nn}
\end{figure}

We deploy a relatively simple, physics-inspired architecture shown in Figure \ref{fig:nn} called \cannonball\ (\textit{Combinatorial Artificial Neural Network ON a BAckronym Lorentz Layer}) based on an idea proposed for top-tagging \cite{Butter:2017cot,Erdmann:2018shi,RPThesis} that a so-called \emph{Lorentz Layer} can help efficiently utilize the relationships between four-vectors. Events are input as matrices with each row representing a separate four-momentum $(E,\vec{p})$. A \emph{Combination Layer} enables the network to add four-momenta by taking weighted linear combinations of the input using a single learned matrix and softmax activation.\footnote{The softmax ensures that no negative weights are applied while forming the combinations, as those could result in objects with $m^{2}<0$.} The fact that this matrix is learned may provide robustness against merging of nearby objects, although this has not been studied.
In this implementation, a \emph{Lorentz Layer} computes the properties $(m,p_{\mathrm{T}},\eta,\phi)$ of those combinations.\footnote{This differs from Ref.~\cite{Butter:2017cot} as there are no learned weights nor exotic observables computed. Through testing, this switch resulted in faster training and slightly better performance.} Lastly, a \emph{Head} uses a batch normalization and feed-forward network to convert these features into an event label, described below.\footnote{The batch normalization was trained for the first epoch to fix the normalization for the remainder of the training; alternate schemes were tested and resulted in lower performance.}

Simulated signal and background events were used to study performance. These samples were produced using \textsc{MadGraph5\_aMC@NLO} 2.7.3 interfaced to \textsc{MadEvent} and the \textsc{RPVMSSM UFO}~\cite{Alwall:2014hca,Fuks:2012im}. Production of $\tilde{t}$ pairs from $13$~TeV $pp$ collisions was simulated with up to one additional parton in the matrix element from ISR. Each $\tilde{t}$ is forced to decay to two light-flavor quarks. One million events at $\tilde{t}$ mass from $0.3$ to $2$ every $0.1$ TeV were generated and split 75/25\% for training and testing, resulting in a total training set size of 13.5M events. For simplicity, these events are analyzed at parton level as a function of the $\tilde{t}$ mass and a transverse momentum smearing that transforms $p_\mathrm{T} \rightarrow \mathcal{N}(p_\mathrm{T},\epsilon p_\mathrm{T})$ for a smearing parameter $\epsilon$ to account for loss of information from fragmentation, hadronization, and detector effects.\footnote{In a dedicated study, $\epsilon$ values of roughly 20\% captured the spread due to a parton shower model from \textsc{Pythia} 8.2 \cite{Sjostrand:2006za,Sjostrand:2014zea}.}
Using $\epsilon$ rather than looking at reconstruction-level jets enables a quantitative understanding of how the algorithms perform as information is removed. Throughout this paper, only parton-level results are presented where loss of information is represented in a controlled way via $\epsilon$.
The angular extent of the hadronic shower, and therefore the potential for object merging, is ignored in this study. A background sample of QCD events containing exclusively $pp\rightarrow 4/5j$ was also generated at parton level, where $j$ represents a parton.

Since the benchmark data is at parton level with up to one additional parton from ISR, \cannonball's input is a $5\times 4$ event matrix, representing five four-vectors. There are no events with more than five partons, and those with four have the fifth row zero-padded. Explicit masking within the architecture for this padding was tested and did not improve results.  The target event label is an 8-bit sequence; the first five bits identify which 4 partons are produced in the $\tilde{t}$ decays, while the last three detail how to combine them. The network produces a likelihood score for each bit.

In an attempt to achieve signal mass invariance, a single network was trained on all mass points at a given $\epsilon$. During preprocessing, events were shuffled so that the mass points were mixed with roughly even statistics per batch, and the input four-vectors were randomly ordered per event. We also tried $p_{\mathrm{T}}$ ordering the inputs and saw better performance for fewer-epoch trainings but comparable performance for longer trainings. 

The networks were implemented using \textsc{PyTorch} \cite{NEURIPS2019_9015} and trained on an NVIDIA Quadro RTX using \textsc{CUDA} version 11.5 \cite{cuda}. Binary Cross Entropy was minimized using the Adam optimizer over 30 epochs with a learning rate of $10^{-3}$ and a batch size of 10k \cite{kingma2017adam}. The large batch size was important so that each mass sample was well-represented in the training. The \emph{Combination Layer} and \emph{Head} were chosen to have 30 combinations and three 200-node hidden layers, respectively. The number of learned parameters was 110k, corresponding to $\sim 0.8\%$ of the training set size. Networks were also trained for 20 times longer with the same architecture and for the same duration with 10$\times$-wider layers in the Combination Layer and Head, leading to only a $\sim$1\% improvement, so the shorter length and smaller network was favored for reproducibility.

\begin{figure}[t]
    \includegraphics[width=8.6 cm]{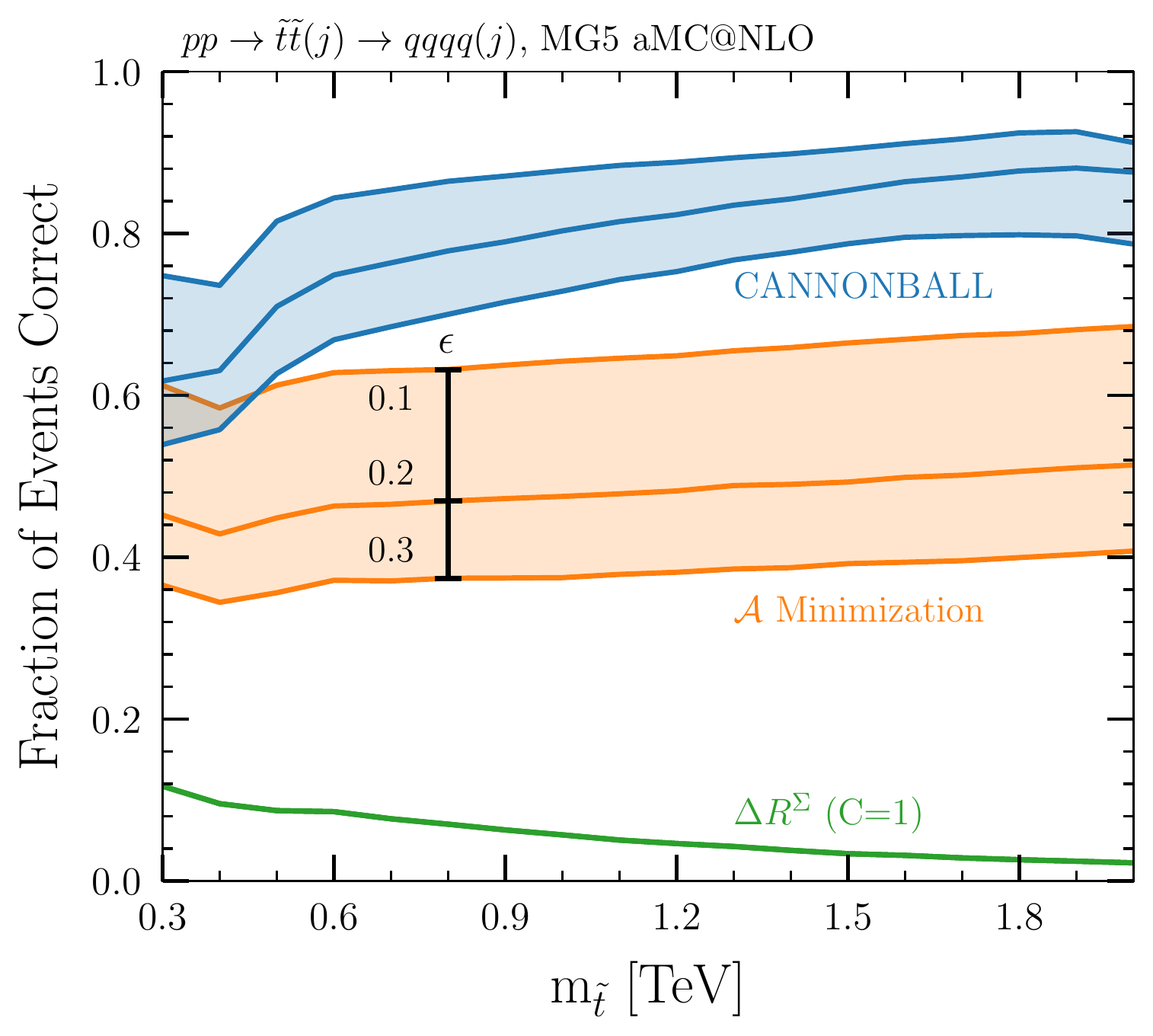}
    \caption{The fraction of correctly reconstructed events for \cannonball\ (blue), $\mathcal{A}$ minimization (orange), and $\Delta R^{\Sigma}$ minimization (green) as a function of stop mass for $p_{\mathrm{T}}$ smearing between $\epsilon=0.1$ and $0.3$.}
    \label{fig:bandplot}
\end{figure}

Figure~\ref{fig:bandplot} shows the fraction of correctly reconstructed events versus stop mass for \cannonball, $\mathcal{A}$ minimization, and $\Delta R^{\Sigma}$ minimization on the testing set. The bands indicate the degree of p$_{\mathrm{T}}$ smearing from $\epsilon=0.1$ to $0.3$. The classical methods are both significantly outperformed by the neural network. The $\Delta R^{\Sigma}$ minimization performance does not change with $\epsilon$ since the smearing does not modify the $\eta - \phi$ distributions and performs the worst because most of the stops are not highly boosted. The accuracy decreases with stop mass because the ratio $\left(2m/p_{\mathrm{T}}\right)_{\tilde{t}}$ increases with mass, resulting in less collimated decay products. \cannonball\ performs significantly better than either of the classical methods across the full range of masses but with more mass dependence than the $\mathcal{A}$ minimization. This is expected since at lower masses the kinematics of the signal partons are more similar to those of ISR, but the general shape is heavily influenced by the training procedure, architecture choices, and the loss of information from nonzero $\epsilon$ values. At $\epsilon=0.3$ and $m_{\tilde{t}}=2$~TeV, \cannonball\ gives twice the performance of the $\mathcal{A}$ minimization, and thirty-five times that of the $\Delta R^{\Sigma}$ minimization. 

\begin{figure}[t!]
    \centering
    \includegraphics[width=8.6 cm]{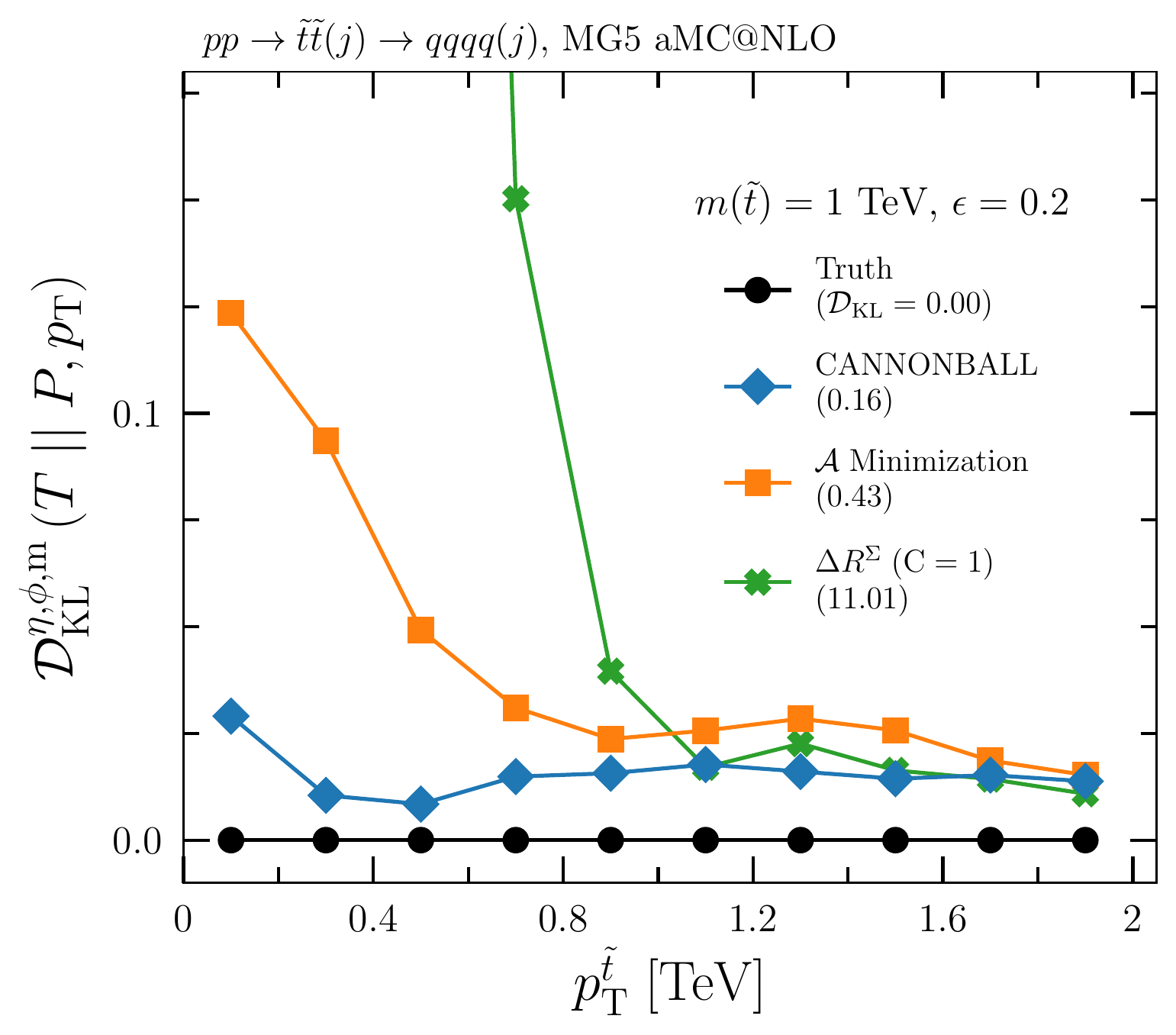}
    \caption{KL divergence over the four-momentum phase space $p^{\mu} = (p_{\mathrm{T}},\eta,\phi,m)$ space in bins of $p_{\mathrm{T}}$ of \cannonball\ (blue), $\mathcal{A}$ minimization (orange), $\Delta R^{\Sigma}$ minimization (green) with respect to truth for $m_{\tilde{t}} = 1$ TeV with $\epsilon=0.2$. Zero divergence (black) is shown for visual aid. The total divergences $\mathcal{D}_{\mathrm{KL}}$ are shown in parentheses.}
    \label{fig:kld}
\end{figure}

While Figure~\ref{fig:bandplot} captures the performance as a function of stop mass, it is useful to understand in which regions of phase space the methods perform well. To quantify this, we calculate the Kullback-Leibler (KL) divergence of the four-dimensional stop momentum probability density functions (PDFs)~\cite{kullback1951information}. The KL divergence is an entropic distance measure between multi-dimensional PDFs defined here as

\begin{align}
    \mathcal{D}_{\mathrm{KL}}\left(T || P\right) &= \int T\log\left(\frac{T}{P}\right)dp^{\mu}   \\
    &= \sum_{p_{\mathrm{T}} \; \mathrm{bins}} \;\sum_{\eta,\phi,\mathrm{m} \; \mathrm{bins}} T\log\left(\frac{T}{P}\right)\\
    &= \sum_{p_{\mathrm{T}} \; \mathrm{bins}} \mathcal{D}_{\mathrm{KL}}^{\eta,\phi,\mathrm{m}}\left(T || P, p_{\mathrm{T}}\right)
\end{align}

\noindent where $p^{\mu} = (p_{\mathrm{T}},\eta,\phi,m)$ is a stop four-momentum and the second line converts $T$ and $P$ from continuous PDFs to $4$D histograms of truth and prediction, respectively. A representative example of $\mathcal{D}_{\mathrm{KL}}^{\eta,\phi,m}\left(T || P, p_{\mathrm{T}}\right)$ is plotted in Figure~\ref{fig:kld} as a function of stop $p_{\mathrm{T}}$ for $m_{\tilde{t}} = 1$~TeV and $\epsilon=0.2$. Smaller values represent more accurate interpretations of the events. The distribution between truth and itself is shown in black as a reference for zero divergence. There is a strong $p_{\mathrm{T}}$ dependence for the classical methods that is not present for the neural network, indicating that the largest performance gains are in the low stop $p_{\mathrm{T}}$ regime. This provides strong evidence that \cannonball\ is best at predicting the full kinematics of the parent stops.

\begin{figure}[t]
    \centering
    \includegraphics[width=8.6 cm]{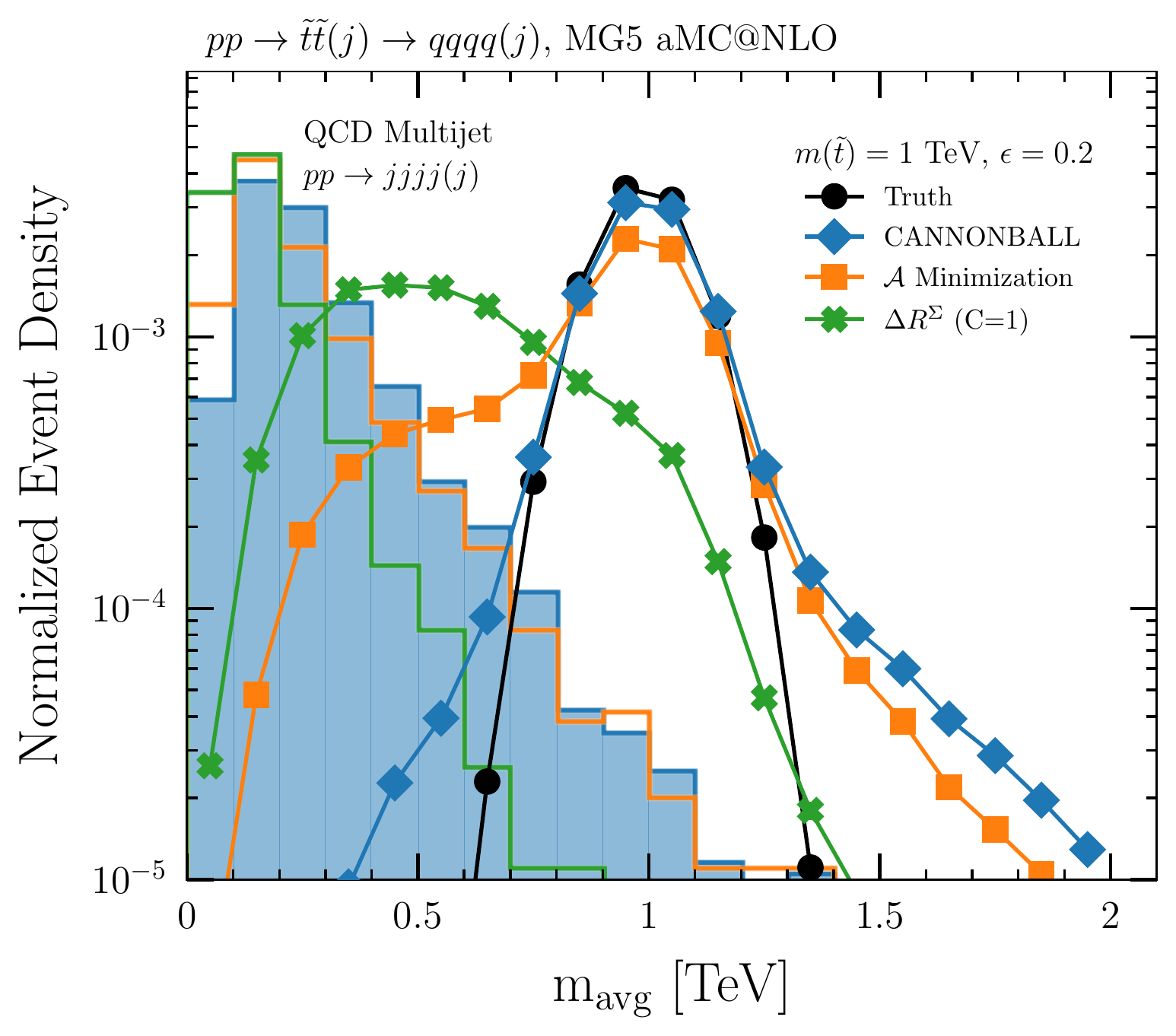}
    \caption{A shape comparison of the per-event average candidate mass from \cannonball\ (blue), $\mathcal{A}$ minimization (orange), $\Delta R^{\Sigma}$ minimization (green) for $m_{\tilde{t}} = 1$ TeV with $\epsilon=0.2$ and simulated QCD background.}
    \label{fig:mavg}
\end{figure}

When using these methods in a bump-hunt search, shape discrimination between the QCD multijet background and the hypothesis signal is crucial. Failing to correctly solve the combinatorial problem leads to decreased discrimination power. In particular, analyses often search in the average mass space $m_{\mathrm{avg}} = \frac{1}{2}\left(m_{1} + m_{2}\right)$, where $m_{1}$ and $m_{2}$ are the predicted candidate stop masses. \cannonball\ results in a more sharply peaked distribution than either of the other methods across all mass points considered. Figure~\ref{fig:mavg} shows a shape comparison between the resulting $m_{\mathrm{avg}}$ distributions for the same signal model shown in Figure~\ref{fig:kld}. \cannonball\ most faithfully reproduces the true signal lineshape.

The blue filled histogram in Figure~\ref{fig:mavg} shows the shape of the QCD background contribution after applying the interpretation given by \cannonball, giving a smoothly falling distribution after an initial kinematic turn-on. None of these three methods significantly sculpt the QCD background distribution away from a steeply falling distribution in the mass range of interest.
Figure~\ref{fig:mavg} also shows the resulting QCD distributions from the classical methods, showing no significant sculpting.
Final discovery sensitivity of each method will depend on the comparison between the shapes from QCD and signal for a given combinatorial solution. Determination of the properties of such a new particle would be best done using \cannonball.

In addition to its ability to improve event reconstruction, this approach would also easily fit into current analysis paradigms. The background estimation techniques, systematic uncertainty evaluation, observable construction, and statistical treatment will still apply from the classical iterations of this analysis. In its simplest form, its just that the separation power of existing variables will be improved by decreasing the contributions from combinatorial backgrounds.\footnote{For this exact signature, a ML approach to constructing an optimal ABCD observable plane is suggested in Ref.~\cite{Kasieczka:2020pil}, although no attempt is made at improving the combinatorial treatment.}

The ability to use classical treatments for the evaluation of systematic uncertainties is a major advantage of this method. In previous work, no dedicated systematic was ever placed on the combinatorial solution method. Instead, experimental uncertainties were propagated to the details of the background estimate and on the final observables~\cite{ATLAS:2017jnp,CMS:2018mts}. Replacing these classical combinatorial solutions with that from \cannonball\ gives no additional source of systematic uncertainty. Existing techniques fully cover the analysis uncertainty, without the complications seen when full-event classification networks are used~\cite{Bhimji:2017qvb,Guo:2018hbv}. 

There are several outstanding questions left for future work. One drawback of this architecture is that it operates via matrix multiplications on full events and therefore requires fixed-size inputs. In signatures with many jets, users would have to increase the architecture size considerably or exclude jets from consideration. A more robust method could natively handle variable size inputs. Theoretical analysis of the architecture, its functional representation and Jacobian, and generalizations are needed to better understand what is learned. Lastly, we aim to explore how to effectively use parton-level information to constrain learned representations of smeared or reconstruction-level events.

In summary, we have identified that the combinatorial background often becomes the primary challenge at high multiplicities and demonstrated a considerable performance increase on a simple $(2\times2)$ multijet signature as a function of resonance mass assuming realistic resolutions. Future work will extend the complexity of the signature and therefore the network, making those problems from well-motivated signatures tractable. More broadly, we advocate for an increased program of ML focusing on kinematic problems where there is an over-abundance, rather than a lack, of information.

We thank Javier Montejo Berlingen, Kate Pachal, Zach Marshall, Steve Farrell, Kelechi Ukah, the Harvard LPPC, Anna Sfyrla and her group, the ATLAS RPV multijet group, and Dan Guest for useful discussions and suggestions. This work has been supported by the Department of Energy, Office of Science, under Grant No. DE-SC0007881 (J.H., A.B., and L.L.), and the Harvard Frederick Sheldon Traveling Fellowship (A.B.). This project has received funding from the European Research Council (ERC) under the European Union's Horizon 2020 research and innovation programme (grant agreement 787331).

\bibliography{MLForMultijets} 

\end{document}